\documentclass[%
reprint,
amsmath,
amssymb,
aps,
]{revtex4-2}

\usepackage[utf8]{inputenc}

\usepackage{graphicx}
\usepackage{cprotect}
\usepackage[export]{adjustbox}

\makeatletter
\let\old@makecaption=\@makecaption
\usepackage{subcaption}
\let\@makecaption=\old@makecaption
\makeatother
\usepackage{dcolumn}
\usepackage{booktabs}
\usepackage{array}
\usepackage{multirow}
\usepackage{tabularx}
\usepackage[inline]{enumitem}

\usepackage{bm}
\usepackage{siunitx}
\DeclareSIUnit\angstrom{\text{Å}}
\usepackage{braket}
\usepackage{mathtools}

\usepackage{amsfonts}
\usepackage{amsthm}

\usepackage{hyperref}
\usepackage{spverbatim}

\usepackage{printlen}
\usepackage{xspace}

\usepackage[capitalize]{cleveref}
\usepackage{xr}

\newcommand{\Rcite}[1]{Ref.~\cite{#1}}

\usepackage{relsize}
\newcommand{\smallunderscore}{\textscale{.5}{\textunderscore}}

\usepackage{derivative}

\makeatletter
\DeclareRobustCommand\onedot{\futurelet\@let@token\@onedot}
\def\@onedot{\ifx\@let@token.\else.\null\fi\xspace}

\def\eg{e.g\onedot}

\def\ie{i.e\onedot}

\def\wrt{w.r.t\onedot}

\makeatother

\newcommand{\QE}{\texttt{Quantum ESPRESSO}\xspace}

\newcommand{\WAN}{\texttt{Wannier90}\xspace}

\newcommand{\AiiDA}{\texttt{AiiDA}\xspace}
\newcommand{\WJL}{\texttt{Wannier.jl}\xspace}

\newcommand{\kpt}{$k$-point\xspace}
\newcommand{\kpts}{\kpt{}s\xspace}
\newcommand{\bvector}{$\mathbf{b}$-vector\xspace}
\newcommand{\bvectors}{\bvector{}s\xspace}

\newcommand{\SrVOthree}{SrVO\textsubscript{3}\xspace}
\newcommand{\MoStwo}{MoS\textsubscript{2}\xspace}

\newcommand{\NatwoSe}{Na\textsubscript{2}Se\xspace}
\newcommand{\CafourOfourteenVfour}{Ca\textsubscript{4}O\textsubscript{14}V\textsubscript{4}}
\newcommand{\HK}{HK}

\newcommand{\epfl}{Theory and Simulations of Materials (THEOS), and National Centre
for Computational Design and Discovery of Novel Materials (MARVEL), \'Ecole
Polytechnique F\'ed\'erale de Lausanne, 1015 Lausanne, Switzerland}
\newcommand{\psich}{Laboratory for Materials Simulations (LMS),
Paul Scherrer Institut (PSI), CH-5232 Villigen PSI, Switzerland}

\begin{document}

\title{Automated mixing of maximally localized Wannier functions into target manifolds}

\author{Junfeng Qiao}
\email{junfeng.qiao@epfl.ch}
\affiliation{\epfl}
\author{Giovanni Pizzi}
\affiliation{\epfl}
\affiliation{\psich}
\author{Nicola Marzari}
\affiliation{\epfl}
\affiliation{\psich}

\date{\today}

\begin{abstract}
  Maximally localized Wannier functions (MLWFs) are widely used to construct
    first-principles tight-binding models that accurately reproduce the electronic
    structure of materials.
  Recently, robust and automated approaches to generate these MLWFs have emerged,
    leading to natural sets of atomic-like orbitals that describe both the occupied
    states and the lowest-lying unoccupied ones (when the latter can be
    meaningfully described by bonding/anti-bonding combinations of localized
    orbitals).
  For many applications, it is important to instead have MLWFs that describe only
    certain target manifolds separated in energy between them---the occupied
    states, the empty states, or certain groups of bands.
  Here, we start from the full set of MLWFs describing simultaneously all the
    target manifolds, and then mix them using a combination of parallel transport
    and maximal localization to construct orthogonal sets of MLWFs that fully and
    only span the desired target submanifolds.
  The algorithm is simple and robust, and it is applied to some paradigmatic but
    non-trivial cases (the valence and conduction bands of silicon, the top valence
    band of \MoStwo, the $3d$ and $t_{2g}$/$e_g$ bands of \SrVOthree) and to a
    mid-throughput study of 77 insulators.
\end{abstract}

\maketitle

\section{Introduction}

Maximally localized Wannier functions (MLWFs)
  \cite{Marzari1997,Souza2001,Marzari2012,Pizzi2020} are accurate reduced-order
  models \cite{Pau2007} for the electronic structures of periodic crystals.
The generation of MLWFs from Bloch wavefunctions typically requires a choice of
  initial projectors, which are often conjectured from chemical intuition with
  trial and error.
For metals, or when considering both the valence and conduction bands (VCB) of
  insulators and semiconductors, one typically deals with bands that overlap
  (\ie, are entangled) \cite{Souza2001} with higher-energy bands.
In such cases, since the low-energy electronic structure can often be well
  described by a tight-binding model of atomic-like orbitals, the initial
  projectors are usually chosen from the hydrogenic $s,p,d,f$ orbitals.
However, when it comes to the case of identifying the optimal projectors for
  valence bands (VB) alone, or especially conduction bands (CB) which are mixed
  with higher-energy bands, it might become difficult to find good initial
  guesses.
Indeed, the VB/CB often consists of bonding/anti-bonding orbitals, or
  combination of atomic orbitals which are more challenging to guess or to
  describe, unless the crystal offers a very simple chemical picture.
Achieving separate Wannierization of target manifolds is also advantageous for
  many applications.
Some physical properties (such as the electric polarization) depend only on the
  Wannier functions (WFs) of the occupied manifold (sum of Wannier centers of all
  the valence WFs).
Moreover, using dedicated MLWFs means that one can obtain smaller tight-binding
  models that are thus more efficient when computing, \eg, transport properties
  of large systems.
Koopmans spectral functionals also require separate occupied and unoccupied
  manifolds \cite{DeGennaro2022}.
Last, low-energy models, such as those used in correlated-electrons
  calculations \cite{Georges1996,Kotliar2006,Maier2005,Gull2011}, require a
  description of the correlated manifold.

Several approaches have been developed in the past few years to simplify the
  construction of MLWFs.
The selected columns of the density matrix (SCDM) algorithm \cite{Damle2015}
  uses QR decomposition with column pivoting on the density matrix to
  automatically generate initial projection orbitals, and a sensible choice of
  the density matrix can be obtained from the projectability of Bloch states onto
  pseudo-atomic orbitals from pseudopotentials \cite{Vitale2020}.
The optimal projection functions method \cite{Mustafa2015} starts with a larger
  manifold and generates the MLWFs of the valence manifold by a single rotation
  matrix, which is computed by a product of a series of Givens rotations.
The dually localized Wannier functions method \cite{Mahler2022} adds an
  additional term to spread functional, to localize the WFs in both space and
  energy, achieving a separation of VB and CB.

Here, we propose a different approach to automatically mix optimal MLWFs
  spanning valence and conduction into several submanifolds, provided that these
  submanifolds are gapped in their energy spectrum.
This naturally applies to the case of separate Wannierizations of valence and
  conduction manifolds, but more generally extends to arbitrary groups of bands
  separated in energy.
We start from the Wannierization of a larger manifold (\eg, the VCB manifold),
  that we do not discuss here since robust methods already exist: in addition to
  hydrogenic $s,p,d,f$ initial projections, partly occupied WF method
  \cite{Thygesen2005,Thygesen2005a}, the fully automated SCDM method
  \cite{Damle2015,Vitale2020}, or the projectability-disentangled Wannier
  function (PDWF) that we recently introduced are available (in particular, the
  latter appears as a very general and remarkably robust approach allowing, \eg,
  to construct $\sim 1.3$ million PDWFs for $\sim 22$ thousands materials
  \cite{PDWF}).
Once these MLWFs are obtained, we then diagonalize the Wannier Hamiltonian at
  every \kpt and partition the states into submanifolds (\eg, valence,
  conduction): they are grouped together if they fall inside the desired energy
  interval.
Next, we fix the gauge randomness of the submanifolds using parallel transport
  \cite{Gontier2019}.
Finally, the MLWFs for each submanifold are generated by maximally localizing
  their spread functionals, independently.
Since the submanifolds are already isolated in energy (\ie, no disentanglement
  is needed), and parallel transport provides a continuous gauge, the final
  maximal localization converges effortlessly.
In the case of separating VB and CB, the final two groups of MLWFs span the
  fully occupied valence and the fully unoccupied conduction manifolds, and their
  shapes closely resemble bonding and anti-bonding orbitals, respectively.
Compared with SCDM, the present method works fully in reciprocal space,
  reducing the computer memory requirements and also being computationally
  faster.
Compared with the optimal projection functions method \cite{Mustafa2015} or the
  dually localized Wannier functions method \cite{Mahler2022}, we do not change
  the spread functional but use the original one in \citet{Marzari1997}, thus the
  resulting WFs are maximally-localized in their original definition; moreover,
  the parallel transport step is non-iterative and always quickly provides a good
  starting point for the final maximally-localization step, avoiding potential
  convergence issues that might occur in an iterative method.

In the following, we first discuss and validate the present method, which we
  name manifold-remixed Wannier function (MRWF), on the VCB of a 3D material
  (silicon), the VCB of a 2D material (\MoStwo), the top VB of \MoStwo, and the
  $3d$ manifold of \SrVOthree.
We also discuss the bonding/anti-bonding character of the resulting MLWFs, as
  well as band interpolation accuracy.
To analyze statistics of band interpolation quality and demonstrate the
  robustness of the present approach, we Wannierize the VB and CB of a diverse
  set of 77 insulators, with the number of atoms between 1 and 45.

\section{Results}

\subsection{The manifold separation algorithm}

While obtaining the starting WFs is not the focus of this paper, we remind here
  that the standard Wannierization algorithm \cite{Marzari1997,Souza2001}
  requires initial projection orbitals $\ket{g_{n}}$ to guide the spread
  minimization and find the most meaningful minimum and the related unitary
  transformation matrices $U_{\mathbf{k}}$ at each \kpt $\mathbf{k}$.
The projectors $\ket{g_{n}}$ are used to rotate the original Bloch
  wavefunctions $\ket{\psi_{m \mathbf{k}}}$ into
  \begin{equation}
    \label{eq:wan_init_proj} \ket{\tilde{\psi}_{n \mathbf{k}}} = \sum_{m=1}^{M}
    \ket{\psi_{m \mathbf{k}}} \braket{\psi_{m \mathbf{k}} | g_{n}},
  \end{equation}
  where $n$ and $m$ are the indices for WFs and Bloch bands, respectively; $M$ is
  the total number of Bloch bands; and $\mathbf{k}$ is the Bloch quasi-momentum.
Note that $\ket{\tilde{\psi}}$ are independent of any arbitrary rotation gauge
  for the $\ket{\psi_{n\mathbf{k}}}$.
For metals or for VCB of insulators, one typically starts with hydrogenic
  $s,p,d,f$ orbitals \cite{Marzari1997} as the initial guesses for all the
  corresponding valence electrons.
Then, the MLWFs can be generated using either the standard disentanglement
  \cite{Souza2001} and maximal localization algorithms \cite{Marzari1997} or
  minimizing directly the total spread, such as the partly occupied WF method
  \cite{Thygesen2005} or a variational formulation \cite{Damle2019}.
Instead of hydrogenic orbitals, one can use SCDM \cite{Damle2015} or the
  recently introduced projectability disentanglement \cite{PDWF} for a fully
  automated Wannierization.
Irrespective of the approach taken to obtain MLWFs describing the VCB, these
  MLWFs will be the starting point of the present algorithm, with the next step
  to separate \eg the VB and CB manifolds from the disentangled MLWFs that span
  both simultaneously.
Note that while in the following we use the separation of VB and CB as an
  example to illustrate the method for clarity and simplicity, the present
  approach is not limited to the case of two submanifolds, but can be applied to
  any groups of bands separated in energy.

Since the disentanglement procedure aims at obtaining the lowest-possible
  spreads, it typically achieves this goal by mixing states originating from all
  the submanifolds (\eg, VB and CB) of interest.
To decompose the manifold into two orthogonal submanifolds, we diagonalize the
  Wannier-gauge Hamiltonian ${H}^{W}_{\mathbf{k}}$ (the superscript $W$ indicates
  the Wannier gauge),
  \begin{equation}
    \label{eq:wan_ham_diag} H^{W}_{\mathbf{k}}
    = V_{\mathbf{k}} \mathcal{E}_{\mathbf{k}} V^{*}_{\mathbf{k}},
  \end{equation}
  where $\mathcal{E}_{\mathbf{k}}$ and $V_{\mathbf{k}}$ are the eigenvalues and
  the eigenvectors, respectively; $*$ denotes conjugate transpose.
Usually the eigenvalues and eigenvectors returned from linear algebra computer
  programs are already sorted in ascending order of eigenvalues; if not, we sort
  them in ascending order, so that the matrices are partitioned into two
  blocks,
  \begin{equation}
    \begin{aligned}
      \mathcal{E}_{\mathbf{k}} & =
      \begin{bmatrix}
        \mathcal{E}^1_{\mathbf{k}} & 0 \\
        0 & \mathcal{E}^2_{\mathbf{k}}
      \end{bmatrix}, \\
      V_{\mathbf{k}} & =
      \begin{bmatrix}
        V^1_{\mathbf{k}} & V^2_{\mathbf{k}}
      \end{bmatrix}, \\
    \end{aligned}
  \end{equation}
  where
  $V^1_{\mathbf{k}} \in \mathbb{C}^{N \times P}$ 
  ($V^2_{\mathbf{k}} \in \mathbb{C}^{N \times Q}$) corresponds to states
  whose eigenvalues $\mathcal{E}^1_{\mathbf{k}}$ ($\mathcal{E}^2_{\mathbf{k}}$)
  are below (above) the band gap, and $0$ represents a zero matrix.
Here, $N$ is the number of WFs of the VCB manifold, $P$ and $Q$ are the number
  of WFs in the valence (below band gap) and the conduction (above band gap)
  submanifolds, respectively, such that $N = P + Q$.
Next, all the Wannier-gauge operators are rotated according to
  $V^1_{\mathbf{k}}$ for the valence submanifold: for instance, the overlap
  matrices $M^{W}_{\mathbf{k}, \mathbf{b}}$ (for computing the spread functional)
  is rotated by
  \begin{equation}
    \label{eq:wan_gauge_rot}
    M_{\mathbf{k}, \mathbf{b}}^{W,1}
    = V^{1*}_{\mathbf{k}} M^{W}_{\mathbf{k}, \mathbf{b}} V^1_{\mathbf{k+b}},
  \end{equation}
  where
  \begin{equation}
    \begin{aligned}
      M^{W}_{\mathbf{k}, \mathbf{b}} & 
      = U^*_{\mathbf{k}} M_{\mathbf{k}, \mathbf{b}} U_{\mathbf{k} + \mathbf{b}}, \\
      M_{\mathbf{k}\mathbf{b}} & 
      = \braket{ u_{m,\mathbf{k}} | u_{n,\mathbf{k+b}} },
    \end{aligned}
  \end{equation}
  $U_{\mathbf{k}}$ are the unitary transformations from the VCB manifold
  Wannierization, and $\ket{u_{m,\mathbf{k}}}$ is the periodic part of Bloch
  wavefunction
  $\ket{\psi_{m,\mathbf{k}}} = e^{i \mathbf{k}\mathbf{r}} \ket{u_{m,\mathbf{k}}}$.
For more details on the notations of $M_{\mathbf{k}, \mathbf{b}}$ and
  $\mathbf{b}$-vectors, see \citet{Marzari1997}.
Consistently, the $\mathcal{E}^1_{\mathbf{k}}$ is used as the new eigenvalues.
Now the problem is reformulated into a Wannierization of an isolated
  submanifold with $P$ WFs for VB.
Similarly, the conduction manifold operators are rotated by $V^2_{\mathbf{k}}$,
  and an analogous Wannierization of an isolated submanifold with $Q$ WFs.
Indeed, the first-step Wannierization of VCB has already disentangled the MLWFs
  from all the remaining higher-energy bands, so that these MLWFs span all the
  target submanifolds we are interested in.
To achieve our goal, we are left with Wannierizations of two isolated
  submanifolds and thus the subsequent steps do not need any disentanglement.
Such a two-step procedure makes the whole algorithm quite robust, especially
  when Wannierizing the CB, for which it is difficult to provide good initial
  projections of the corresponding anti-bonding orbitals.

The remaining difficulty of the Wannierization of the two isolated submanifolds
  is caused by the diagonalization in \cref{eq:wan_ham_diag}.
Indeed, since the Hamiltonians $H^{W}_{\mathbf{k}}$ are independently
  diagonalized at each $\mathbf{k}$, the resulting eigenvectors will have
  different gauges at different \kpts, requiring additional Wannierizations in
  each submanifold.
Since these Wannierizations are carried out on submanifolds that have isolated
  bands, the minimization algorithm is typically more robust to the choice of
  initial projections compared to the case of disentanglement.
One could simply resort to random Gaussian initial projections followed by
  maximal localization to reach the MLWFs for the two submanifolds, respectively;
  or even brute-force maximal localization starting from the random gauge after
  the Wannier Hamiltonian diagonalization.
However, a direct maximal localization starting from a random gauge is not
  robust---we observe that, in many cases, the maximal localization fails due to
  zeros in the diagonal of the overlap matrices $M_{\mathbf{k}, \mathbf{b}}$;
  and, even if it converges, it displays the same issues of random Gaussian
  projections: a large number of iterations, and oscillatory evolution of spread
  and the sum of MLWF centers during the minimization process (see
  \cref{supp-fig:si2_val_4gauge} in supplementary information (SI) for
  discussions on the convergences of these choices).
Moreover, when the number of \kpts $N_{\mathbf{k}}$ is large, the maximal
  localization is much harder to converge.
A better solution is finding good starting gauge for the two submanifolds in an
  automated fashion.

To tackle this challenge, we adopt the parallel transport algorithm
  \cite{Gontier2019} to construct smooth gauges for the two submanifolds.
For an isolated manifold, the existence of a smooth gauge is determined by its
  topological obstructions, which are characterized by the Chern numbers (one in
  2D and three in 3D).
If the Chern numbers are 0 (as it is the case for systems with time-reversal
  symmetry), then it can be proven \cite{Brouder2007} that it is possible to
  construct a continuous gauge explicitly by the following procedure
  \cite{Gontier2019}: (a) Suppose $k_i \in [0, 1]$ (in fractional coordinates)
  for $i=x,y,z$: propagate (using singular value decomposition of overlap
  matrices $M_{\mathbf{k}, \mathbf{b}}^{W}$ to maximally align the gauge between
  neighboring \kpts) the Bloch wavefunctions $\ket{ u_{n\mathbf{0}} }$ at
  $\Gamma$ along $k_x$ from \kpt $\mathbf{0} = (0, 0, 0)$ to $\mathbf{1} = (1, 0,
    0)$, to construct a continuous gauge across these \kpts.
The new gauge is not necessarily quasi-periodic, \ie, satisfying the Bloch
  theorem imposed on $\ket{ u_{n\mathbf{k}} }$ by $\ket{ u_{n\mathbf{k+K}} } =
    \tau_{\mathbf{K}} \ket{ u_{n\mathbf{k}} }$ where $\tau_{\mathbf{K}}=e^{-i
    \mathbf{K} \cdot \mathbf{r}}$ is the translation operator in reciprocal space,
  and $\mathbf{K}$ is a reciprocal lattice vector.
In general, instead, the two states are related by $\ket{ u_{n\mathbf{1}} } = (
    \tau_{\mathbf{1}} \ket{ u_{n\mathbf{0}} } ) V_{\mathrm{obs}}$, where this
  expression defines the obstruction matrix $V_{\mathrm{obs}}$ quantifying the
  misalignment of the propagated gauge and the gauge required by Bloch theorem at
  $(1,0,0)$.
To fulfill the quasi-periodic boundary condition, we can therefore multiply
  each $\ket{ u_{n\mathbf{k}} }$ by $e^{-k_x L}$ (note $k_x \in [0,1]$) where
  $V_{\mathrm{obs}} = \exp(L)$: in this way, we obtain a continuous gauge that
  also satisfies Bloch theorem, \ie, the new obstruction matrix in this modified
  gauge is the identity matrix.
(b) For each $k_x$, propagate along $k_y$ from $(k_x, 0, 0)$ to $(k_x, 1, 0)$.
Now we obtain a series of obstruction matrices $V_{\mathrm{obs}}(k_x)$ along
  $k_x$.
If the winding number \cite{Cances2017,Cornean2017a,Gontier2019} of the
  determinants of $V_{\mathrm{obs}}(k_x)$ vanishes (\ie, the Chern number is 0),
  then there is a continuous function that maps $V_{\mathrm{obs}}(k_x)$ to
  identity \cite{Gontier2019}.
We then multiply the gauge by this mapping, so that the new gauge satisfies the
  quasi-periodic boundary condition in the $k_x-k_y$ plane.
\Rcite{Gontier2019} explicitly constructs the continuous mapping by their
  ``column interpolation'' method for the Kane-Mele model, which is a 2D
  fermionic time-reversal-symmetric model (\ie, having a vanishing Chern number)
  but can present a non-zero $\mathbb{Z}_2$ number; as a comparison, previous
  methods had difficulties in handling $\mathbb{Z}_2$ systems
  \cite{Damle2017,Cornean2016,Gontier2019}, sometimes requiring model-specific
  information \cite{Mustafa2016,Winkler2016}.
(c) For each ($k_x, k_y$), propagate along $k_z$ from $(k_x, k_y,
  0)$ to $(k_x, k_y, 1)$.
Now the obstruction matrices $V_{\mathrm{obs}}(k_x, k_y)$ depend both on $k_x$
  and $k_y$.
Similar to point (b), if the two winding numbers of the determinants of
  $V_{\mathrm{obs}}(k_x, 0)$ and $V_{\mathrm{obs}}(0, k_y)$ vanish, then there is
  a continuous function that maps $V_{\mathrm{obs}}(k_x, k_y)$ to identity.
We then multiply the gauge with this mapping and obtain the final gauge
  satisfying the quasi-periodic boundary condition in 3D.
\Rcite{Gontier2019} demonstrates this constructive algorithm to obtain a
  continuous gauge for a 3D system (silicon).
The results also show that the continuous gauge can be further smoothened by
  the standard maximal localization procedure \cite{Marzari1997} to construct
  MLWFs.
We stress that the algorithm is non-iterative and fast, thus solving the
  problem of finding good initial WFs for isolated manifolds in an efficient and
  robust way.

As shown in \cref{supp-fig:si2_val_4gauge}, parallel transport generates a much
  better starting point than random Gaussian projections or random gauges: the
  convergence of maximal localization is much faster, and the spread and the sum
  of MLWF centers smoothly evolve during minimization.
We note that since the propagation of gauge requires overlap matrices between a
  particular set of nearest-neighboring \kpts, in the SI \cref{supp-sec:bvec_pt}
  we present a procedure so that parallel transport can be applied to any
  arbitrary crystal structure.

In summary, the sequential parallel transports move the obstructions to the
  Brillouin zone edges, and the ``column interpolation'' method fixes the
  quasi-periodicity.
Our tests on a set of 77 insulators (see discussion later in
  \cref{sec:77materials}) show that this algorithm is able to construct a good
  initial gauge, and maximal localization is able to construct MLWFs without
  issue.

We now mention that since we propagate the gauge starting from the first \kpt
  $(0,0,0)$, there is still one gauge arbitrariness at this $\Gamma$ point.
Here, we suggest to insert an additional step that first minimizes the spread
  functional \wrt a single rotation matrix $W$ for the first \kpt, before
  performing the standard maximal localization \wrt all \kpts to obtain MLWFs.
Indeed, thanks to the small size of $W$, this first preliminary step is
  computationally efficient, and can help in further improving the overall
  robustness of the full algorithm that we are presenting here.
To achieve this, we derive the expression of the gradient of the spread
  $\Omega$ \wrt the rotation matrix $W$ in SI \cref{supp-sec:opt_rot}.
We then use this gradient with a manifold optimization algorithm
  \cite{Mogensen2018} to minimize $\Omega$ \wrt $W$, where $W$ is constrained
  on the unitary matrix manifold
  $\left\{ W \in \mathbb{C}^{K \times K} \vert W^{*} W = I \right\}$,
  where $K = P$ for the valence manifold, or $K = Q$ for the conduction manifold.
This minimization provides us with a single rotation matrix $W$ that further
  improves the localization, while still preserving the parallel transport gauge:
  we stress that, in addition to increasing the robustness of the algorithm as
  mentioned earlier, this additional step can thus be beneficial for cases where
  the parallel transport gauge is implicitly assumed during the derivation of
  equations (for instance, Wannier interpolation of Berry curvature
  \cite{Wang2006}, or Wannier interpolation of nonlinear optical responses
  \cite{Wang2017,IbanezAzpiroz2018}).

After the parallel transport and the single rotation, the resulting WFs are
  close to the ideal MLWFs.
However, since parallel transport only generates a continuous quasi-periodic
  gauge, it typically does not provide the smallest possible spread.
It is therefore helpful to perform a final smoothing of the gauge
  \cite{Gontier2019} by running a final maximal localization step (see examples
  in \cref{sec:silicon,sec:mos2}).
This can be achieved using either the original Marzari-Vanderbilt localization
  \cite{Marzari1997} or a matrix manifold optimization \wrt gauge matrices
  at all the \kpts, \ie, optimization on a product manifold of a series of
  unitary matrices 
  $\prod_{\mathbf{k}} \left\{ U_{\mathbf{k}} \in \mathbb{C}^{K \times K}
  | U^*_{\mathbf{k}} U_{\mathbf{k}} = I\right\}$,
  where $K=P$ for valence and $K=Q$ for conduction manifolds.
As already mentioned, the multi-step procedure that we propose here aims at
  making the whole algorithm more robust, since every step produce a better
  starting point for the final iterative localization algorithm.

In summary, we start from an initial manifold that has been already singled out
  from the remaining high-energy states using standard procedure such as
  disentanglement and maximal localization (\eg, very accurately using
  projectability disentanglement \cite{PDWF} to extract as much as possible the
  bonding and anti-bonding characters from all the bands).
The subsequent diagonalizations of Wannier-gauge Hamiltonians separate the
  manifold into (two) orthogonal submanifolds (for VB \& CB, respectively).
The (two) parallel-transport steps (for the relevant submanifolds) construct
  continuous gauges, fixing the randomness caused by the independent Hamiltonian
  diagonalization at each \kpt.
The rotation \wrt a single unitary matrix removes the gauge arbitrariness of
  parallel transport at the first \kpt.
The final maximal localizations ultimately smoothen the gauge, leading to two
  sets of MLWFs, each of which spans the submanifold for VB or CB.
In SI \cref{supp-sec:MRWF_gauge}, we prove that the final gauge transformation
  has block diagonal structure, \ie, the MRWFs are transformed according to
  \begin{equation} \label{eq:mrwf_gauge}
    U(\mathbf{k}) =
    \begin{bmatrix}
      U_{\mathrm{VB}}(\mathbf{k}) & 0 \\
      0 & U_{\mathrm{CB}}(\mathbf{k})
    \end{bmatrix},
  \end{equation}
  where $U_{\mathrm{VB}}(\mathbf{k})$ and $U_{\mathrm{CB}}(\mathbf{k})$ are
  unitary matrices for VB and semi-unitary matrices for CB, respectively.

\subsection{Silicon} \label{sec:silicon}
\begin{figure}[!htb]
  \includegraphics[width=\linewidth,max height=0.6\textheight]{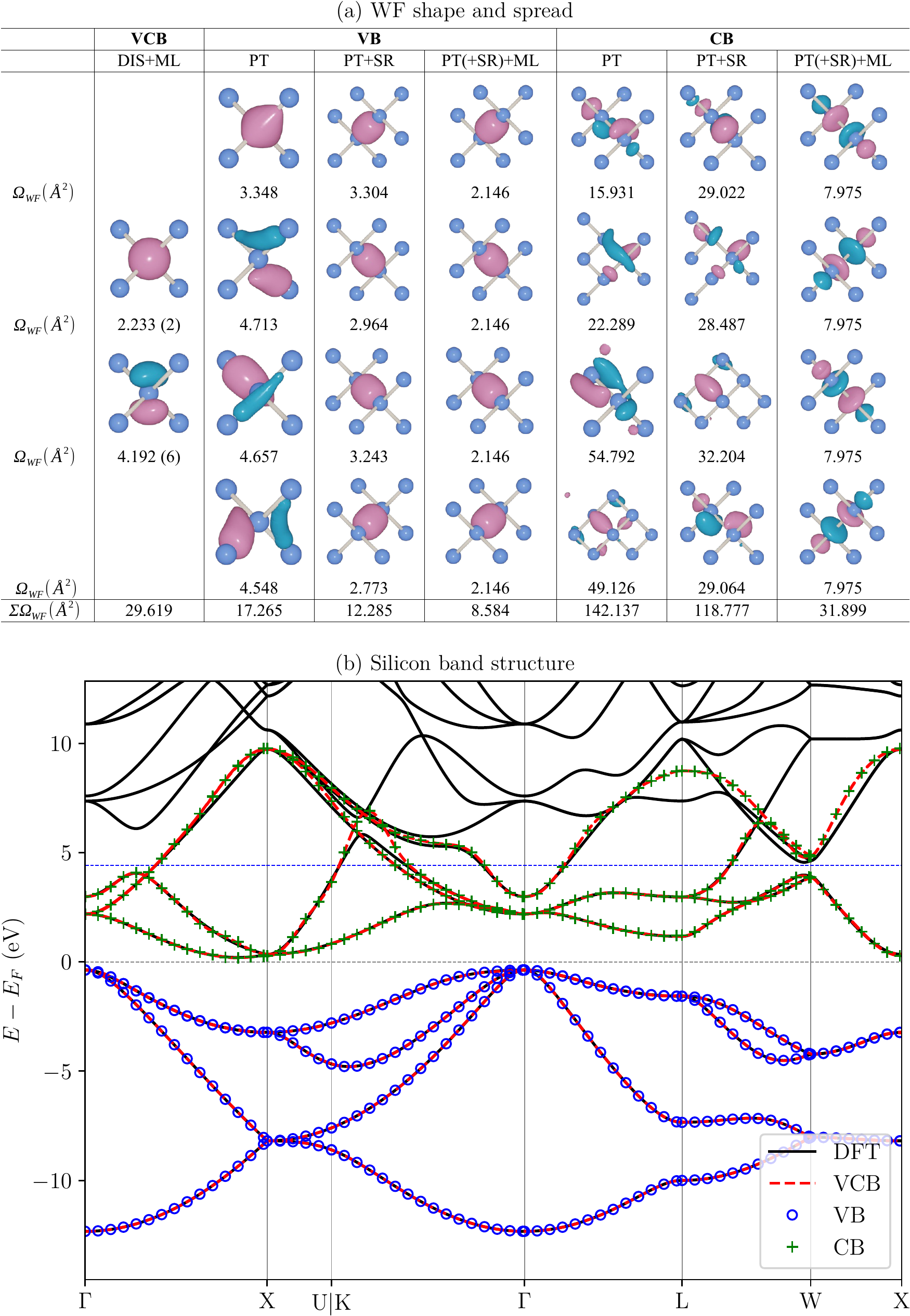}
  \begin{subcaptiongroup}
    \phantomsubcaption\label{fig:Si2&shape}
    \phantomsubcaption\label{fig:Si2&band}
  \end{subcaptiongroup}
  \cprotect\caption{\textbf{WF shapes and band interpolation of silicon.}
    \subref{fig:Si2&shape} WF shape of various Wannierization methods.
    In the header, the left, center, and right columns correspond to the valence
      plus conduction bands (VCB), valence bands (VB), and conduction bands (CB),
      respectively.
    The PT column shows WFs after running parallel transport (PT); The PT+SR column
      shows WFs after running PT and single rotation (SR) of $W$ matrix; the
      PT(+SR)+ML column shows WFs after running PT (and optionally SR) and maximal
      localization (ML), since PT+SR+ML and PT+ML fall to the same minimum.
    The numbers under each WF are the spread and its multiplicity: \eg, 1.668 (2)
      means there are two WFs having spread \SI{1.668}{\angstrom^2}.
    The last row $\sum \Omega_{\mathrm{WF}}$ is the total spread of each
      calculation.
    The blue spheres are the silicon atoms.
    \subref{fig:Si2&band} Band structure comparison of DFT, VCB, VB, and CB.
    The horizontal dashed blue line is the upper limit of the frozen window for
      disentanglement (DIS).
    The VCB are computed by Wannier interpolation from DIS+ML; the VB and CB are
      computed by Wannier interpolation from PT+ML.
  } \label{fig:Si2}
\end{figure}

To test the validity of the present method, we first disentangle and maximally
  localize the VCB of silicon into 8 WFs, using the standard hydrogenic $s$ and
  $p$ projections with energy window disentanglement (we use hydrogenic
  projectors and energy window disentanglement here to demonstrate that the
  present approach works well as long as the entire VCB are accurately described;
  one can also use PDWF to construct MLWFs spanning the entire VCB).
The resulting WFs, two of which have $s$-character and six of which have
  $p$-character, are shown in the VCB column of \cref{fig:Si2&shape}.

For the valence manifold, after running the Hamiltonian diagonalization and
  parallel transport, we obtain four WFs with spreads around \SI{4}{\angstrom^2},
  but their shapes do not have clear physical meaning, since the gauge of the
  first \kpt is still arbitrary (see PT column inside the VB column of
  \cref{fig:Si2&shape}); after running the single rotation of the $W$ matrix,
  their spreads are further minimized to around \SI{3}{\angstrom^2}, and the
  shapes now resemble the bonding orbitals between neighboring silicon atoms
  (PT+SR column inside the VB column of \cref{fig:Si2&shape}); a final maximal
  localization further reduces the spreads to around \SI{2}{\angstrom^2} and the
  four spreads end up becoming identical (PT(+SR)+ML column inside the VB column
  of \cref{fig:Si2&shape}) thus respecting the symmetry of the full system.
For the conduction manifold (CB columns of \cref{fig:Si2&shape}), again the
  shapes of WFs after parallel transport have no clear meaning.
However, even after the single rotation, the shapes of WFs still do not
  resemble the expected anti-bonding orbitals, and only after the final maximal
  localization the anti-bonding shape is recovered.
Note that, in this simple case of silicon, for both valence and conduction
  manifolds we reach the same set of WFs whether we run a maximal localization
  directly after parallel transport, or a maximal localization after parallel
  transport + single rotation, so the two cases are merged into one column in
  \cref{fig:Si2&shape} under the header PT(+SR)+ML.
The total spreads for the VB and the CB manifolds after parallel transport +
  maximal localization are \SI{8.584}{\angstrom^2} and \SI{31.899}{\angstrom^2},
  respectively.
As expected, their sum is larger than the value for the VCB manifold
  (\SI{29.619}{\angstrom^2}) after disentanglement and maximal localization,
  since in the VCB case there is additional freedom to further minimize the
  spread by remixing bonding and anti-bonding WFs into pure $s$ and $p$ orbitals
  (we highlight that using atom-centered $s,p$ projections does not lead to the
  most localized orbitals for VCB in silicon; with a choice of atom-centered
  $sp^3$ projections, the total spreads can further decrease to
  \SI{26.761}{\angstrom^2}, where four WFs have spreads \SI{3.522}{\angstrom^2}
  and another four \SI{3.168}{\angstrom^2}).
In addition, we note that since the Hamiltonian diagonalization returns a
  random gauge, the spreads for parallel transport and parallel transport +
  single rotation are different in each run, but the spreads of PT(+SR)+ML should
  always be the same, since the algorithm should always manage to find the
  maximally-localized gauge in this simple case.
To quantify how our multi-step procedure increases the overall robustness of
  the algorithm while at the same time reducing its computational cost, we show
  in SI \cref{supp-fig:si2_val_4gauge} the evolution of WF spreads and centers
  during maximal localizations.
Starting from the random gauge directly after Hamiltonian diagonalization
  (\cref{supp-fig:si2_val_4gauge&rand_gauge}), it takes \num{28430} iterations to
  converge; using random Gaussians as initial guesses
  (\cref{supp-fig:si2_val_4gauge&rand_proj}), the number of iterations decreases
  significantly to \num{812}; with the parallel transport gauge
  (\cref{supp-fig:si2_val_4gauge&pt}), the number of iterations further decreases
  to \num{228}, and the evolution of spreads and centers is much smoother; the
  best starting gauge is the one after single rotation
  (\cref{supp-fig:si2_val_4gauge&pt_or}), which only takes \num{40} iterations to
  converge, without any oscillations in the evolution of spreads and centers.
Note that the spreads of valence MRWFs from PT(+SR)+ML are the same as MLWFs
  obtained from a direct Wannierization of the valence bands, \ie, the valence
  MRWFs after separation span the original DFT valence manifold, thus the initial
  VCB Wannierization does not cause any delocalization of the valence MRWFs (see
  SI \cref{supp-sec:MRWF_gauge} for a proof).

We now discuss the quality of the band interpolation.
The WFs for VB \& CB are constructed from the initial VCB manifold obtained
  from a preliminary disentanglement and maximal localization.
Therefore, if VB/CB are properly Wannierized, their band interpolation quality
  should be similar to that of VCB MLWFs.
Thus, in the following paragraphs, we compare the band interpolations of VB/CB
  MLWFs, VCB MLWFs, and DFT bands.
Once the starting VCB manifold was properly disentangled and could well
  reproduce the DFT band structure, as shown in \cref{fig:Si2&band}, the WFs
  after parallel transport + maximal localization for the VB and the CB manifolds
  can reproduce the corresponding part of the VCB Wannier interpolated bands with
  high accuracy, being visually indistinguishable.
To quantitatively evaluate the band interpolation quality, we compute the
  average band distance, $\eta_{\mathrm{isolated}}$, between the VCB and VB/CB
  bands \cite{Prandini2018,Vitale2020,PDWF}.
The $\eta_{\mathrm{isolated}}$ is defined as
  \begin{equation}
    \label{eq:bandsdist} \eta_{\mathrm{isolated}}^{\mathrm{A-B}} = \sqrt{\frac
      {\sum_{n\mathbf{k}} (\epsilon_{n\mathbf{k}}^{\mathrm{A}} -
        \epsilon_{n\mathbf{k}}^{\mathrm{B}})^2 } {N_{b} N_{\mathbf{k}}} },
  \end{equation}
  where $\epsilon_{n\mathbf{k}}$ are the eigenvalues of a band
  structure, and its superscript A or B refers to the eigenvalues of two
  different bands, A or B, which can be DFT bands, or Wannier interpolated bands
  of VCB, VB, or CB; $N_b$ and $N_{\mathbf{k}}$ are the number of bands and
  \kpts, respectively.
For silicon, we obtain: $\eta_{\mathrm{isolated}}^{\mathrm{VCB-VB}} =$
  \SI{6.6}{meV} and $\eta_{\mathrm{isolated}}^{\mathrm{VCB-CB}} =$
  \SI{15.5}{meV}.
In general, the VB interpolation is more accurate than CB since the VB MLWFs
  usually have smaller spreads.
To improve the CB interpolation quality, one might need to increase \kpt
  sampling, as we discuss in \cref{sec:mos2_topVB}.

\subsection{\MoStwo}\label{sec:mos2}

Next, we test the method on a two-dimensional (2D) \MoStwo monolayer.
For VCB Wannierization, we use the standard hydrogenic Mo $d$ and S $s,p$
  projections (the semicore states are excluded, so in total 9 VBs and the lowest
  4 CBs are Wannierized).
Since the VB and the lowest four CBs of \MoStwo are isolated, 13 WFs are
  maximally localized from 13 bands without disentanglement.
The 13 MLWFs can be well characterized into 4 groups by their angular momentum:
  as shown in the VCB column of \cref{fig:MoS2&band}, from top to bottom, 3
  resemble $d_{z^2}$, 2 resemble $d_{xy}$, and the remaining 8 resembles $sp^3$
  hybridized orbitals.

\subsubsection{Valence and conduction bands}
For the valence manifold, both after PT and after PT+SR, the WFs still do not
  have a clear resemblance to bonding orbitals; after PT+ML or PT+SR+ML, the WFs
  can be well grouped into six hybrids of Mo $d_{z^2}$ + S $p$, two $s$-like WFs
  near sulfur atoms, and one WF floating inside the hexagonal cage and having
  $C_{3h}$ symmetry, originating from the hybridization of three properly
  oriented Mo $d_{z^2}$ orbitals from the three nearest Mo atom.
For the conduction manifold, WFs after PT are already close to the anti-bonding
  hybrid orbitals, and the further SR or ML steps help to slightly reduce the
  spreads and result in more symmetrized WF shapes.
We notice that in contrast to intuition, in this case it took more iterations
  to converge starting from PT+SR gauge than directly from the PT gauge, as shown
  in the SI \cref{supp-fig:MoS2_evolution}.
Although the PT+SR cases start from a smaller total spread, the maximal
  localizations got stuck longer in plateaus in the final stages, leading to
  longer iterations.
However, in all cases, the evolutions are smooth and converge in less than
  \num{500} iterations, since both valence and conduction are isolated manifolds
  themselves, and PT is able to construct good starting gauges.
\begin{figure}[!htb]
  \includegraphics[width=\linewidth,max height=0.6\textheight]{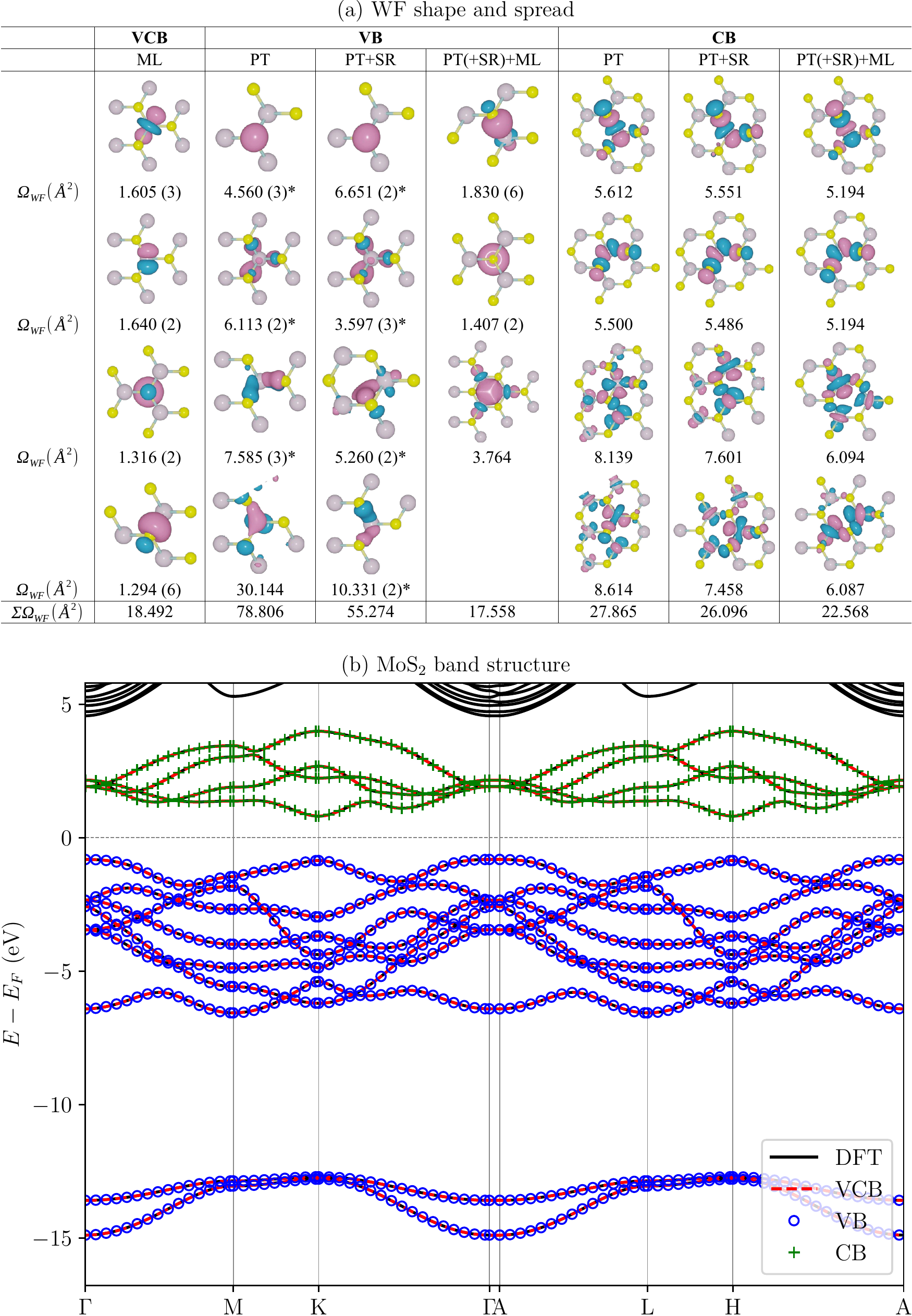}
  \begin{subcaptiongroup}
    \phantomsubcaption\label{fig:MoS2&shape}
    \phantomsubcaption\label{fig:MoS2&band}
  \end{subcaptiongroup}
  \cprotect\caption{\textbf{WF shapes and band interpolation of \MoStwo.}
    \subref{fig:MoS2&shape} WF shape of various Wannierization methods.
    The notations are the same as \cref{fig:Si2&shape}, except that in some cases
      the additional star sign (*) indicates that the WFs are grouped together if
      their spreads are roughly similar, and only one of the shapes is shown: \eg,
      4.560 (3)* means there are three WFs having similar spreads, and their average
      spread is \SI{4.560}{\angstrom^2}.
    The yellow and the silver spheres are the S and Mo atoms, respectively.
    \subref{fig:MoS2&band} Band structure comparison of DFT, VCB, VB, and CB.
    The VCB are computed by Wannier interpolation after maximal localization (ML);
      the VB and CB are computed by Wannier interpolation from PT+ML.
    The DFT bands are almost indistinguishable since the VCB, VB, and CB
      Wannier-interpolated bands overlap essentially exactly with the DFT bands.
  }
  \label{fig:MoS2}
\end{figure}

In terms of band interpolation, again the Wannier interpolated VB and CB
  overlap essentially exactly with the DFT bands as well as the Wannier
  interpolated VCB, as shown in \cref{fig:MoS2&band}, and demonstrated
  quantitatively by the excellent band-distance values:
  $\eta_{\mathrm{isolated}}^{\mathrm{VCB-VB}} =$ \SI{0.19}{meV} and
  $\eta_{\mathrm{isolated}}^{\mathrm{VCB-CB}} =$ \SI{0.51}{meV}.

\subsubsection{Single top valence band}\label{sec:mos2_topVB}

In practical applications, the highest valence and lowest conduction bands are
  of high interest since they are critical for electronic transport properties.
However, the Wannierization of a single band remains elusive since it is
  difficult to write down a proper initial projection, resulting from a complete
  hybridization of many different atomic orbitals.
However, in the \MoStwo case, since the top valence band is isolated with
  respect to all other bands, we can use our algorithm to construct a smooth
  gauge for that single band, demonstrating the more general applicability of our
  method, beyond the separation of VB \& CB.

\begin{figure}[!htb]
  \includegraphics[width=\linewidth,max height=0.6\textheight]{
    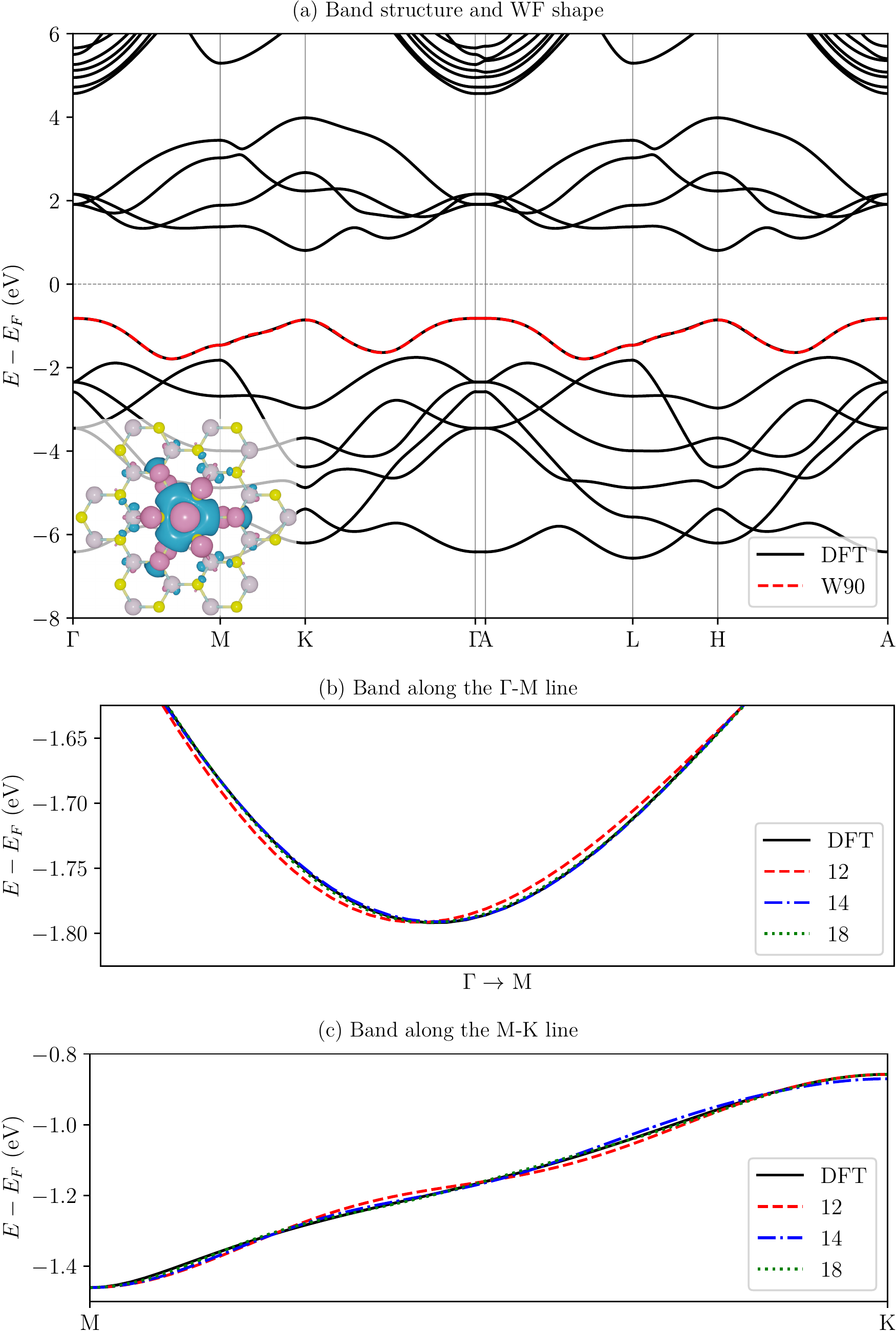}
  \begin{subcaptiongroup}
    \phantomsubcaption\label{fig:MoS2_topVB&band}
    \phantomsubcaption\label{fig:MoS2_topVB&zoom_GM}
    \phantomsubcaption\label{fig:MoS2_topVB&zoom_MK}
  \end{subcaptiongroup}
  \cprotect\caption{\textbf{Band structure comparison
      of the single \MoStwo top valence band.}
    \subref{fig:MoS2_topVB&band} Comparison of DFT and Wannier-interpolated top VB.
    The bottom-left inset shows the shape of the resulting single MLWF.
    \subref{fig:MoS2_topVB&zoom_GM} and \subref{fig:MoS2_topVB&zoom_MK} are zoom-in
      comparisons of the Wannier-interpolated bands using different $k$-point
      samplings: red dashed line for $12 \times 12 \times 1$, blue dotted dashed line
      for $14 \times 14 \times 1$, and green dotted line for $18 \times 18 \times 1$.
    The $16 \times 16 \times 1$ is not shown since it has similar quality to the
      $14 \times 14 \times 1$.
  }
  \label{fig:MoS2_topVB}
\end{figure}

\Cref{fig:MoS2_topVB&band} shows the band interpolation of this single-band WF,
  and the inset shows the shape of this highly symmetric WF in real space.
As usual, since the separate Wannierizations in each submanifold have less
  degrees of freedom compared with the Wannierization of the initial manifold,
  the WF spreads for separate Wannierization are usually larger.
Indeed, the single WF has a relatively large spread (\SI{9.288}{\angstrom^2}).
For such a large spread, artificial interactions between periodic copies of the
  same WF in different supercells (where the supercell size is determined by the
  \kpt sampling) may start to become non-negligible.
Indeed, we observe some small oscillations at the minimum of the band along
  $\Gamma$ to M and along M to K, whose zoom-ins are shown in
  \cref{fig:MoS2_topVB&zoom_GM,fig:MoS2_topVB&zoom_MK}.
By increasing the $k$-point sampling from the $12 \times 12 \times 1$
  (\SI{0.2}{\angstrom^{-1}} density, same as the VCB Wannierization) to $18
    \times 18 \times 1$, the interpolation quality improves significantly (see
  \cref{fig:MoS2_topVB&zoom_GM,fig:MoS2_topVB&zoom_MK}).
This means that the band interpolation error is not caused by our separation
  method, but by the insufficient $k$-point sampling.
Therefore, if one targets a very high band interpolation quality, the $k$-point
  sampling might need to be increased when considering a submanifold only.

\subsection{\SrVOthree}
\begin{figure}[!htb]
  \includegraphics[width=\linewidth,max height=0.6\textheight]{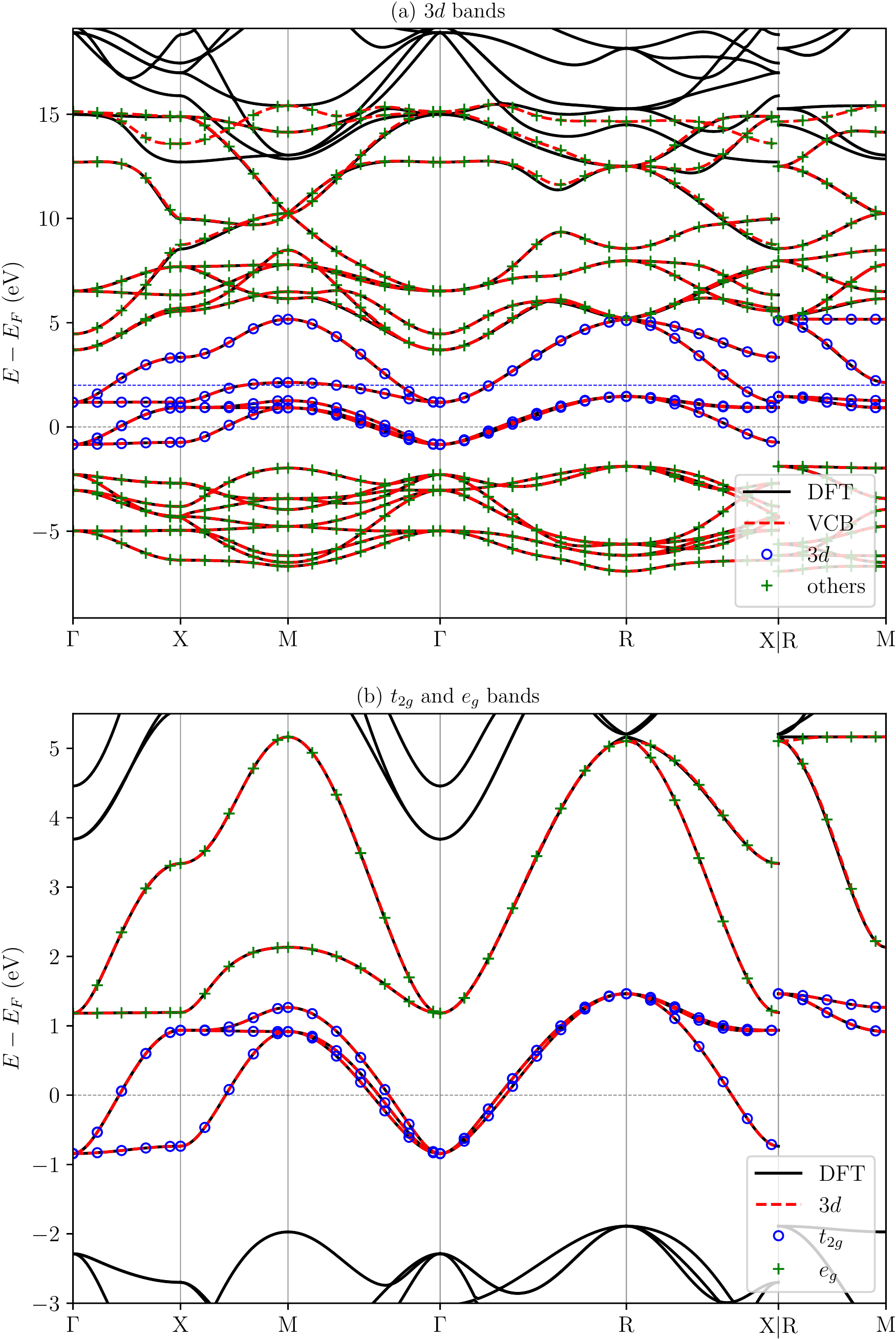}
  \begin{subcaptiongroup}
    \phantomsubcaption\label{fig:SrVO3&band_3d}
    \phantomsubcaption\label{fig:SrVO3&band_t2g_eg}
  \end{subcaptiongroup}
  \cprotect\caption{\textbf{Comparison of DFT and Wannier-interpolated bands for \SrVOthree.}
    \subref{fig:SrVO3&band_3d} The VCB are computed by Wannier interpolation using
      PDWF; the $3d$ bands ($3d$) and the remaining bands (others) are computed by
      Wannier interpolations from PT+ML.
    At DFT level, the $3d$ bands have a small gap (\SI{37.8}{meV} at R point)
      separated from the bands above; the present algorithm successfully separates
      the $3d$ submanifold from the remaining manifold.
    \subref{fig:SrVO3&band_t2g_eg} The $t_{2g}$ and $e_g$ bands are further
      separated starting from the $3d$ bands of \subref{fig:SrVO3&band_3d}.
  }
  \label{fig:SrVO3_band}
\end{figure}

Here we test the present method on the metallic perovskite \SrVOthree, which is
  a correlated material on the $t_{2g}$ Hubbard manifold.
At the DFT level, there is a small gap (\SI{37.8}{meV} at R point) separating
  the $3d$ manifold from bands above, thus satisfying the requirement of isolated
  manifold for the present method.
Starting from 24 MLWFs for VCB, we generate two sets of MLWFs: 5 for $3d$ and
  19 for the remaining manifold.
For brevity, we only show the comparison of DFT and Wannier-interpolated bands
  in \cref{fig:SrVO3&band_3d}, while the respective WF shapes and spreads are
  shown in SI \cref{supp-fig:SrVO3_wf,supp-fig:SrVO3_evolution}.
For the $3d$ manifold, the PT gauge is already quite close to the maximally
  localized gauge: maximal localization only slightly decreases the total spread
  from \SI{9.815}{\angstrom^2} of PT to \SI{9.629}{\angstrom^2}, and symmetrizes
  the shapes of WFs (the $3d$ columns of \cref{supp-fig:SrVO3_wf}).
For the remaining manifold, it is quite hard to converge: only after
  \num{13027} iterations (\cref{supp-fig:SrVO3_evolution}) the maximal
  localization can converge to real-valued, spatially-symmetrized MLWFs
  (\cref{supp-fig:SrVO3_wf}).
In this case, the single rotation greatly helps in improving the convergence:
  only \num{1544} iterations are needed to converge to the same MLWFs starting
  from PT+SR gauge, also removing the oscillations in spreads and centers during
  maximal localization (\cref{supp-fig:SrVO3_evolution}).
For the band interpolation, again the respective manifolds are accurately
  reproduced, as demonstrated by the bands in \cref{fig:SrVO3_band} and the band
  distances: $\eta_{\mathrm{isolated}}^{\mathrm{VCB-}3d} =$ \SI{6.74}{meV} and
  $\eta_{\mathrm{isolated}}^{\mathrm{VCB-others}} = $ \SI{3.00}{meV}.
Furthermore, since the $t_{2g}$ and $e_g$ manifolds are gapped in energy, we
  can also separate them into two submanifolds.
As shown in \cref{fig:SrVO3&band_t2g_eg}, the $t_{2g}$ and $e_g$ bands are
  again reproduced very well.

\subsection{Results on 77 insulators} \label{sec:77materials}
Finally, we test our method on a set of 77 insulators with number of atoms in
  the unit cell ranging from 1 to 45.
This is the same as the insulator set of \Rcite{Vitale2020}, except that 4 (He,
  Ne, Ar\textsubscript{2}, Kr\textsubscript{2}) of the 81 materials are excluded
  since they consist of closed-shell noble-gas atoms, where the valence electrons
  are fully occupied (and there is thus no need for separate Wannierizations).
This comprehensive test set not only validates the correctness of the present
  method, but also helps improve its generality to cover edge cases (\eg, the
  additional treatment of \bvectors in the SI \cref{supp-sec:bvec_pt}) that would
  be difficult to discover with only a few test cases.
The separate Wannierization is implemented as a fully automated \AiiDA
  \cite{Pizzi2016,Huber2020,Uhrin2021} workflow, which first runs the
  Wannierization of VCB using PDWF \cite{PDWF}, then splits the VCB manifold with
  the method discussed here (see \cref{sec:code} for the \WJL code
  implementation), and then runs two separate maximal localizations using \WAN
  for the VB and the CB manifolds, respectively.

\begin{figure}[!htb]
  \includegraphics[width=\linewidth,max height=0.6\textheight]{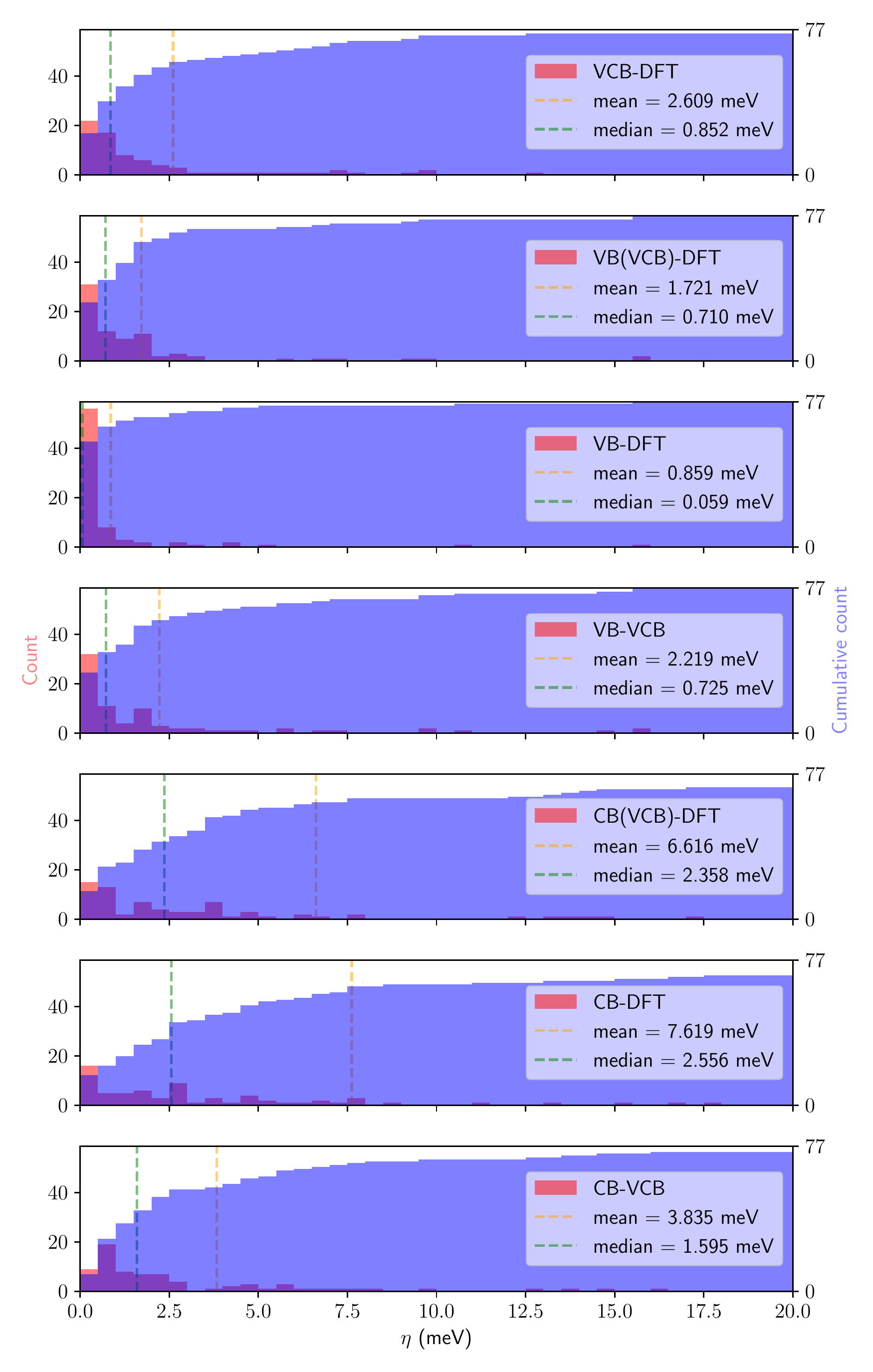}
  \cprotect\caption{\textbf{Band distances of 77 insulators.}
    From top to bottom: histograms of band distances for valence plus conduction
      bands (VCB) \wrt DFT, valence bands (VB) of VCB Wannierization \wrt DFT, VB
      Wannierization \wrt DFT, VB Wannierization \wrt VCB Wannierization, conduction
      bands (CB) of VCB Wannierization \wrt DFT, CB Wannierization \wrt DFT, and CB
      Wannierization \wrt VCB Wannierization.
    The red and blue bars are the histograms and cumulative histograms,
      respectively.
    The vertical lines indicate the mean and median values.
  }
  \label{fig:bandsdist}
\end{figure}

All the Wannierizations finish successfully and have excellent band
  interpolation quality, which we measure by the band distance
  \cite{Prandini2018,Vitale2020,PDWF} $\eta_{\mathrm{isolated}}$ for comparisons
  between isolated bands (VB of separate Wannierization \wrt VB of DFT, VB of
  separate Wannierization \wrt VB of VCB Wannierization, CB of separate
  Wannierization \wrt CB of VCB Wannierization), and $\eta_2$ for comparisons
  involving CB of DFT (VCB Wannierization \wrt VCB of DFT, CB of separate
  Wannierization \wrt CB of DFT),
  \begin{equation}
    \label{eq:eta2}
    \eta_{2}^{\mathrm{A-B}} = \sqrt{\frac {\sum_{n\mathbf{k}}
    \tilde{f}_{n\mathbf{k}} (\epsilon_{n\mathbf{k}}^{\mathrm{A}} -
    \epsilon_{n\mathbf{k}}^{\mathrm{B}})^2 } {\sum_{n\mathbf{k}}
    \tilde{f}_{n\mathbf{k}}} },
  \end{equation}
  where $\tilde{f}_{n\mathbf{k}} =
    \sqrt{f^{\mathrm{A}}_{n\mathbf{k}}(E_{F}+2, \sigma)
      f^{\mathrm{B}}_{n\mathbf{k}}(E_{F}+2, \sigma)}$ and $f(E_{F}+2, \sigma)$ is the
  Fermi-Dirac distribution with the Fermi energy set to \SI{2}{eV} above the real
  Fermi energy $E_F$ to compare also part of the conduction bands; the smearing
  width is set to $\sigma=$ \SI{0.1}{eV}.
The statistics of $\eta$ are shown in \cref{fig:bandsdist}.
It is worth noting that the VB interpolation
  ($\eta_{\mathrm{isolated}}^{\mathrm{VB-DFT}}$ = \SI{0.859}{meV}) is even more
  accurate than the VCB interpolation ($\eta_{2}^{\mathrm{VCB-DFT}}$ =
  \SI{2.609}{meV}): this is partly because $\eta_{2}^{\mathrm{VCB-DFT}}$ is
  averaged over all bands, including the larger error of CB interpolation; to
  exclude the effect of averaging, we also compute the band distance of the VB of
  VCB Wannierization \wrt the VB of DFT,
  $\eta_{\mathrm{isolated}}^{\mathrm{VB(VCB)-DFT}}$ = \SI{1.721}{meV}, which is
  larger than $\eta_{\mathrm{isolated}}^{\mathrm{VB-DFT}}$ = \SI{0.859}{meV},
  showing that the VB interpolation is indeed more accurate than the VCB
  interpolation---this can be explained by two facts: (1) the valence MRWFs are
  constructed by unitary transformations of Bloch states, thus the valence
  manifold is exactly preserved (see SI \cref{supp-sec:MRWF_gauge} for a proof);
  (2) the valence MRWFs are more localized than the VCB MLWFs (will be discussed
  in the next paragraph), leading to a better Fourier interpolation quality.
The CB interpolation ($\eta_{2}^{\mathrm{CB-DFT}}$ = \SI{7.619}{meV}) is
  slightly worse than the CB of VCB Wannierization
  ($\eta_{2}^{\mathrm{CB(VCB)-DFT}}$ = \SI{6.616}{meV}), since the CB MLWFs are
  more delocalized than the VCB MLWFs (will be discussed in the next paragraph);
  moreover, it appears much larger than the $\eta_{2}^{\mathrm{VCB-DFT}}$ =
  \SI{2.609}{meV} since $\eta$ is defined as an average over all bands---the
  (accurate) VB interpolations are not taken into account in the computation of
  $\eta_{2}^{\mathrm{CB-DFT}}$.
In addition, and most importantly, the VB/CB to VCB distances are
  $\eta_{\mathrm{isolated}}^{\mathrm{VB-VCB}}$ = \SI{2.219}{meV} and
  $\eta_{\mathrm{isolated}}^{\mathrm{CB-VCB}}$ = \SI{3.835}{meV}, showing that
  the submanifolds are well separated with little loss of interpolation accuracy
  compared with the starting-point VCB Wannierization.
For completeness, we also show the statistics of max band distance, which is a
  stricter measure of band interpolation quality, in the SI
  \cref{supp-fig:maxbandsdist}.

\begin{figure}[!htb]
  \includegraphics[width=\linewidth,max height=0.6\textheight]{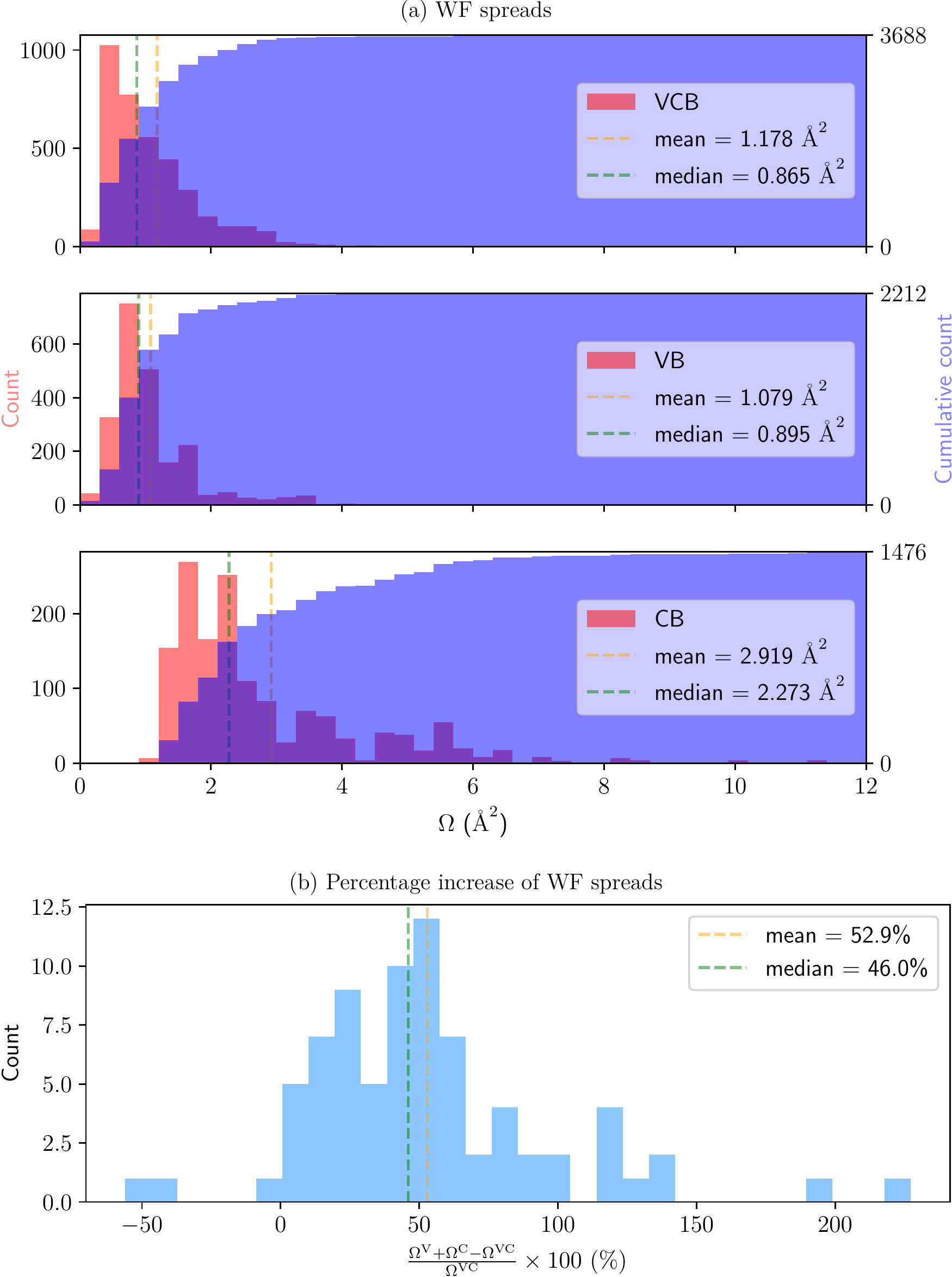}
  \begin{subcaptiongroup}
    \phantomsubcaption\label{fig:spread&hist}
    \phantomsubcaption\label{fig:spread&incr}
  \end{subcaptiongroup}
  \cprotect\caption{\textbf{WF spreads of 77 insulators.}
    \subref{fig:spread&hist} From top to bottom: histograms of VCB, VB, and CB MLWF
      spreads.
    The upper limit of the right y-axis of each panel is the total number of WFs,
      respectively.
    \subref{fig:spread&incr} Histogram of the percentage increase of the sum of VB
      and CB spreads \wrt VCB spreads.
  }
  \label{fig:spread}
\end{figure}

Now we discuss the localization of MLWFs by comparing the average spread from
  the 77 materials.
For VCB Wannierization, the average spread $\Omega^{\mathrm{VCB}}$ =
  \SI{1.178}{\angstrom^2}; after separation (followed by maximal localization of
  VB and CB, respectively), the average spread of VB MLWFs is slightly more
  localized ($\Omega^{\mathrm{VB}}$ = \SI{1.079}{\angstrom^2}); while that of CB
  MLWFs are more delocalized ($\Omega^{\mathrm{CB}}$ = \SI{2.919}{\angstrom^2}).
This is consistent with the intuition that the VB MLWFs are the more localized
  bonding orbitals whereas the CB MLWFs are the more delocalized anti-bonding
  orbitals.
Finally, as discussed in the previous section, the separated MLWFs have less
  degrees of freedom compared with the VCB MLWFs, thus the sum of the spreads of
  VB and CB ($\Omega^{\mathrm{VB}} + \Omega^{\mathrm{CB}}$) is in general larger
  than that of VCB ($\Omega^{\mathrm{VCB}}$).
\Cref{fig:spread&incr} shows the percentage increase of $\Omega^{\mathrm{VB}} +
    \Omega^{\mathrm{CB}}$ over $\Omega^{\mathrm{VCB}}$.
On average, there is a 52.9\% increase of the spread.
Note, however, that in three cases (\NatwoSe, \CafourOfourteenVfour, and \HK)
  there are a 39.2\%, 56.0\%, and 5.1\% decreases of spreads, respectively.
For the first two cases, there are few large-spread VCB MLWFs, probably because
  the VCB Wannierizations are trapped in local minima; during the separation
  Wannierizations, the parallel transport algorithm is able to find a smoother
  gauge, thus reaching more localized MLWFs for both VB and CB, leading to
  smaller $\Omega^{\mathrm{VB}} + \Omega^{\mathrm{CB}}$ than
  $\Omega^{\mathrm{VCB}}$.

For completeness, we show the band-structure comparisons, the band distances,
  the evolution of spreads and the evolution of the sum of WF centers during
  maximal localization in SI \cite{supp}, for each of the 77 materials.
The smooth evolution of spreads and sum of WF centers during maximal
  localization for CB and VB demonstrate that parallel transport is able to
  construct continuous gauge, thus maximal localization has no difficulty in
  further smoothening the gauge.

\section{Conclusions}
We introduce an automated method (manifold-remixed Wannier functions (MRWF)) to
  separate band manifolds by constructing MLWFs for the respective submanifolds
  that have finite energy gaps between them.
The method naturally extends to the case of valence and conduction manifolds,
  but also to any other case of isolated groups of bands.
First, we start with a properly Wannierized valence plus conduction manifold,
  obtained using any manual or automated method (the recently introduced
  projectability disentangled WF \cite{PDWF} is particularly suitable for this
  application since it is able to robustly and reliably construct MLWFs that
  preserve as much as possible the anti-bonding characters).
Then, we split the manifolds by diagonalizing the Wannier-gauge Hamiltonian
  into submanifolds for target energy ranges, respectively.
Next, using parallel transport, we construct smooth gauges for each submanifold
  to fix the randomness caused by the independent Hamiltonian diagonalization at
  every \kpt.
Finally, we maximally localize the parallel-transport gauge to obtain smooth
  MLWFs for the desired manifolds.
Before the final maximal localization, we can optionally run a preliminary
  rotation \wrt a single unitary matrix to fix the remaining gauge randomness
  intrinsic to parallel transport.
Often the final maximal localization is able to find the maximally-localized
  gauge directly; however, the single rotation step helps to improve the
  robustness of the final maximal localization, and has the additional benefit of
  improving localization while still preserving the parallel transport gauge,
  which might be relevant in some applications.

Results on silicon and \MoStwo show that the final valence (conduction) MLWFs
  restore faithfully chemical intuition for bonding/anti-bonding orbitals, and
  accurately reproduce the valence/conduction part of the band structure of the
  valence plus conduction manifold.
Moreover, we demonstrate that the method is not limited to the separation of
  valence and conduction manifolds, but also applicable to any system with band
  groups separated by a finite gap: for instance, the single top valence band of
  \MoStwo; or the $3d$, $t_{2g}$, and $e_g$ manifolds of \SrVOthree.
Furthermore, we implement fully automated \AiiDA
  \cite{Pizzi2016,Huber2020,Uhrin2021} workflows to carry out the whole separate
  Wannierization process, and test the present method on a set of 77 insulators.
Statistics show that the band interpolation achieves excellent accuracy at the
  \si{meV} scale, and on average the sum of VB and CB MLWF spreads increase
  around 50\% \wrt the VCB MLWF spreads.
Thus, we highlight that to ensure accurate band interpolation quality, the \kpt
  sampling density for separated Wannierization might need to be increased, as
  demonstrated in the Wannierization of the \MoStwo top valence band.

As an outlook, we envision several applications that one may find useful with
  the preset approach: the analysis of bonding/anti-bonding orbitals based on
  MLWFs; material properties that rely solely on the occupied manifold, such as
  the electric polarization; spectral theories that require separate sets of
  localized orbitals for both occupied and unoccupied states (for instance, the
  Koopmans functionals to predict accurately the electronic band gap
  \cite{DeGennaro2022}); and the dynamical mean field theory for correlated
  electrons.

\section{Methods} \label{sec:methods}

The DFT calculations are carried out by \QE\cite{Giannozzi2020a}, using the
  SSSP efficiency (version 1.1, PBE functional) library\cite{Prandini2018} for
  pseudopotentials and its recommended energy cutoffs.
The high-throughput calculations for 77 insulators are managed by the \AiiDA
  \cite{Pizzi2016,Huber2020,Uhrin2021} infrastructure which submits \QE and \WAN
  \cite{Pizzi2020} calculations to remote clusters, parses, and stores the
  results into a database, while also orchestrating all sequences of simulations
  and workflows.
The automated \AiiDA workflows are open-source and hosted on \texttt{GitHub}
  \cite{aiidaW90}.
Semicore states from pseudopotentials are excluded from Wannierizations, except
  for a few cases where the semicore states overlap with valence states; in such
  cases, all the semicore states are Wannierized.
A regular $k-$point mesh is used for the Wannier calculations, with a $k-$point
  spacing of \SI{0.2}{\per\angstrom}, as selected by the protocol in
  \citet{Vitale2020}.
Figures are generated by \texttt{matplotlib} \cite{Hunter2007}.

\section{Data Availability}
All data generated for this work can be obtained from the Materials Cloud
  Archive (\url{https://doi.org/10.24435/materialscloud:2f-hs}).

\section{Code Availability} \label{sec:code}
All codes used for this work are open-source; the latest stable versions can be
  downloaded at \url{http://www.wannier.org/} for \WAN,
  \url{https://www.quantum-espresso.org/} for \QE, \url{https://www.aiida.net/}
  for \AiiDA, and \url{https://github.com/aiidateam/aiida-wannier90-workflows}
  for \texttt{aiida-wannier90-workflows}.

The MRWF method is implemented in an open-source \texttt{Julia}
  \cite{Bezanson2017} package named \WJL, which is available at
  \url{https://github.com/qiaojunfeng/Wannier.jl}, and
  \url{https://www.wannierjl.org/} for the accompanying documentation/tutorials.

\section{Acknowledgements}
We thank Antoine~Levitt and Michael~F.~Herbst for helpful discussions and
  feedback on the \WJL implementation.
We acknowledge financial support from the NCCR MARVEL (a National Centre of
  Competence in Research, funded by the Swiss National Science Foundation, grant
  No.
205602), the Swiss National Science Foundation (SNSF) Project Funding
(grant 200021E\smallunderscore{}206190 ``FISH4DIET'').
The work is also supported by a pilot access grant from the Swiss National
  Supercomputing Centre (CSCS) on the Swiss share of the LUMI system under
  project ID ``PILOT MC EPFL-NM 01'', a CHRONOS grant from the CSCS on the Swiss
  share of the LUMI system under project ID ``REGULAR MC EPFL-NM 02'', and a
  grant from the CSCS under project ID s0178.

\section{Author Contributions}
J.~Q.
implemented and tested the method.
G.~P.
and N.~M.
supervised the project.
All authors analyzed the results and contributed to writing the manuscript.

\section{Competing Interests}
The authors declare that there are no competing interests.

\bibliography{main}

\end{document}